\documentclass[notitlepage, 11pt]{article}
\usepackage[margin=1in]{geometry}
\usepackage{amsmath, amssymb, amsthm, mathtools, bm}
\usepackage{titling}
\usepackage{booktabs,siunitx}
\usepackage{graphicx,float,caption,subcaption}
\usepackage[colorlinks=true, linkcolor=blue, citecolor=blue, urlcolor=blue]{hyperref}
\usepackage{fancyhdr}
\usepackage{tabularx}

\usepackage[authoryear, round]{natbib}

\DeclareMathOperator{\logit}{logit}

\begin{document}
	\title{%
		A Latent-Space Statistical Learning Framework for Semicontinuous Outcomes: \\
		Regularized Deviance and Boundary-Stabilized Optimization
	}
	
	\author{Jianping Philip Wang \\ Acuity, A Mutual Insurance Company\thanks{Acuity Insurance, Actuarial and Strategic Analytics; email: philip.wang@acuity.com}}
	
	\maketitle
	
	\begin{abstract}
		Semicontinuous response variables characterized by an exact point mass at zero and heavily right-skewed positive continuous distributions pose persistent estimation challenges across empirical risk modeling, econometrics, and quantitative analytics. Standard single-index formulations (e.g., compound Poisson--Gamma Tweedie processes) impose a rigid mean-variance coupling that distorts the zero-mass probability when confronted with extreme tail volatility, flattening the continuous density and compromising extreme-risk discrimination. Furthermore, heuristic post-hoc target clamping introduces systematic bias while leaving loss gradients vulnerable to tail leverage. 
		
		We present a unified, statistically coherent likelihood framework for semicontinuous modeling under arbitrary zero-inflation and explicit upper boundary constraints. All density formulations, likelihood evaluations, and curvature derivations operate strictly within a latent coordinate space defined by the power-scaling mapping $y_i^* = \min\{(y_i/s)^\lambda, U^*\}$, where $s \in (0, U)$ anchors the distribution and $\lambda \in (0,1)$ regularizes tail curvature. The architecture formulates semicontinuous outcomes as a coherent, dual-target learning task defined over a single shared covariate matrix $\mathbf{X}$, partitioning the response into three operational regimes: (1) a discrete zero-mass governed by a Bernoulli occurrence layer, (2) a finite continuous positive body modeled via an analytical Gamma density, and (3) an upper boundary point-mass governed by lower incomplete Gamma survival ratios. Rather than training disconnected frequency and severity pipelines over fragmented subsets of data, our approach optimizes both outputs concurrently under a single unified deviance objective.
		
		Rather than treating the capping threshold $U$ as an arbitrary truncation, the boundary functions as an intrinsic tail accumulator that bounds the score and Hessian dynamics during optimization. We derive the exact closed-form deviance objective, score vectors, and block-diagonal Hessian matrix for this joint architecture. We prove that the occurrence and severity parameter spaces are information-orthogonal ($\partial^2 \mathcal{D} / \partial \zeta_i \partial \eta_i \equiv 0$), eliminating cross-target gradient contamination while preserving joint covariate representation across both predictive heads. Asymptotic boundary analysis demonstrates that the score and curvature scale linearly in the deep tail, establishing an intrinsic quadratic restoring force that prevents gradient collapse and numerical blowouts. Empirical comparisons illustrate that the proposed framework breaks the Tweedie parameter rigidity paradox, maintaining absolute structural fidelity across both zero occurrence and high-severity risk differentiation.
	\end{abstract}

	\section{Introduction} \label{sec:intro}
	
	Semicontinuous response variables arise naturally across a diverse spectrum of quantitative disciplines, including healthcare econometrics, climatological monitoring, operational risk management, and reliability engineering \citep{Frees2010}. These outcomes are structurally characterized by a mixture of exact discrete mass points at zero and positive continuous values, typically accompanied by substantial right-skewness and extreme tail behaviors \citep{Cragg1971}. Developing coherent statistical models for such distributions remains a fundamental challenge, as their joint empirical characteristics routinely violate the assumptions underlying classical exponential dispersion families and standard regression frameworks.
	
	A distinguishing feature of semicontinuous data is the simultaneous presence of three distinct, often conflicting, factors driving distributional mismatch:
	\begin{enumerate}
		\item The existence of a degenerate point mass at zero, reflecting the absolute absence of an event, exposure, or expenditure.
		\item The severe structural mismatch of highly skewed empirical distributions with familiar, un-regularized theoretical density functions across the positive continuous support.
		\item The presence of extreme right tails whose support may span several orders of magnitude, exerting disproportionate leverage during parameter estimation.
	\end{enumerate}
	
	To mitigate the influence of extreme observations and control tail variance, practitioners in applied fields routinely transform, trim, cap, winsorize, or censor outcome observations to reshape the empirical support before estimation \citep{Boxcox1964}. In many real-world modeling applications, large observations are systematically capped at a deterministic upper threshold $U$ to control leverage. Such preprocessing implicitly alters the information space available to the estimator. When an extreme observation is intentionally represented through a capping mechanism at threshold $U$, the resulting data point no longer conveys that the underlying realization equals $U$; rather, it conveys that the realization is bounded below by $U$, operating as a genuinely point mass observation. 
	
	Nevertheless, conventional statistical software and machine learning objective functions typically continue to treat these preprocessed, capped observations as exact realizations. This standard treatment introduces an un-regularized information mismatch between the objective likelihood being optimized and the boundary conditions actually contained in the data. The present work resolves this foundational inconsistency through the explicit incorporation of a right-capped survival function directly into a unified semicontinuous deviance loss architecture.
	
	Historically, this modeling impasse has been handled via two contrasting paradigms, 
	both of which introduce structural compromises. The first relies on the conventional 
	frequency--severity framework \citep{Klugman2004}. While flexible, this approach bifurcates 
	the problem into two disconnected modeling pipelines: an occurrence model trained on 
	the full portfolio and a conditional severity model trained strictly on the filtered 
	positive subset ($y_i > 0$). This fragmentation inevitably creates discordant feature 
	spaces, inconsistent variable selections, and contradictory risk ordering when the two 
	outputs are multiplied post-hoc. 
	
	The second paradigm relies on the compound Poisson--Gamma Tweedie family of distributions 
	\citep{Jorgensen1987}. The Tweedie model avoids feature fragmentation by optimizing a single 
	unified index over a shared feature matrix, preserving the familiar mechanics of 
	Generalized Linear Models (GLMs) \citep{Dunn2018}. However, as we demonstrate mathematically, 
	coupled single-index architectures suffer from severe parameter rigidity: when confronted 
	with heavy positive tails, the single joint dispersion parameter $\phi$ is forced to absorb 
	extreme tail variance, which mathematically distorts the zero point-mass and flattens the 
	estimated density across the continuous modal body. 
	
	The framework proposed here resolves this foundational trade-off. It retains the principal 
	advantage of Tweedie—a coherent, shared covariate feature space $\mathbf{X}$ optimized under 
	a unified scalar deviance—while providing the distributional freedom of two-part models via 
	information-orthogonal parameter regimes that eliminate cross-target distortion.
	
	To address distributional mismatch between empirical realizations and the theoretical positive density support, this paper formulates the estimation problem in a regularized latent space. Let $y_i \ge 0$ denote the raw observed outcome and $U$ the operational capping threshold. We define the power-scaling coordinate transformation:
	\begin{equation}
		y_i^* = \min\left\{\left(\frac{y_i}{s}\right)^\lambda,  \; U^*\right\}, \quad U^* = \left(\frac{U}{s}\right)^\lambda, 
		\label{eq:latent_mapping}
	\end{equation}
	where $U$ is the fixed capping or right-censoring boundary, $s \in (0, U)$ denotes a scale-preserving reference factor, and $\lambda \in (0, 1)$ governs the curvature of power transformation:
	\begin{itemize}
		\item large positive observations ($y_i > s$) are compressed;
		\item small positive observations ($y_i < s$) are expanded;
		\item threshold positive observations ($ y_i = s$) are unchanged regardless of $\lambda$ values.
	\end{itemize} 
	By mapping both the observed outcomes and the upper capping boundary into the latent coordinates $(y_i^*, U^*)$, the transformation simultaneously compacts the extreme right tail and stabilizes the numerical support prior to density estimation.
	
	This perspective naturally separates the two principal challenges of semicontinuous modeling into a zero-mass problem and a positive-tail problem. Rather than forcing a single coupled distribution on the entire response, the proposed framework treats zero occurrence and regularized positive magnitude as distinct, orthogonal statistical tasks within a unified likelihood structure. We derive the exact closed-form deviance, gradient vector, and block-diagonal Hessian matrix for this joint architecture, equipping it with a quadratic gradient safeguard that entirely eliminates the vanishing gradient pathologies plaguing conventional infinite-series or sigmoidal loss formulations in deep right tails.

	\section{Literature Review and Theoretical Positioning} \label{sec:lit_review}
	
	\subsection{Explicitly Contrast Against Legacy Hurdle Models}
	The decomposition of semicontinuous processes into independent event-occurrence and conditional-magnitude regimes originates from the seminal work of \cite{Cragg1971} on limited dependent variables. By relaxing the rigid structural constraints of classical Tobit formulations \citep{Tobin1958}—which erroneously treat structural zeros as left-censored latent variables—hurdle and two-part models allow the zero-mass probability $p_{i} = \Pr(Y_{i} = 0)$ and the conditional density $f(Y_{i} \mid Y_{i} > 0)$ to be governed by entirely separate parameter spaces. This paradigm has inspired numerous mixture variants, including zero-inflated Gamma, zero-adjusted Gamma, and zero-inflated Lognormal architectures.
	
	While granting considerable parametric flexibility, traditional hurdle models remain fundamentally unequipped to handle directional boundary constraints on the continuous support. When observations are capped or right-censored at an upper threshold $U$, evaluating the joint likelihood requires computing complex continuous cumulative distribution functions to represent the survival mass $\Pr(Y_{i} \geq U)$. In modern high-dimensional or machine learning applications, the lack of analytically tractable gradients with respect to these censored / capped boundaries causes severe numerical instability, forcing practitioners to erroneously treat censored endpoints as exact realizations, completely ignoring the cap generated point mass.
	
	Beyond boundary treatment, an even more fundamental deficiency of conventional hurdle 
	and two-part implementations in modern machine learning is feature-space divergence. 
	Because the occurrence classifier is trained across the full dataset while the severity 
	regressor is fitted strictly on the non-zero subset ($y_i > 0$), tree-splitting algorithms 
	and variable selection routines branch along entirely different feature trajectories. 
	This operational disconnect frequently generates contradictory risk indications: an 
	individual risk profile $\mathbf{x}_i$ may be segmented into a low-risk decile by the 
	occurrence model while simultaneously assigned extreme risk weight by the conditional 
	severity model due to disjoint interaction splits. What is missing in the literature 
	is an architecture that enforces a unified, coherent feature representation $\mathbf{X}$ 
	across the entire sample while preserving the mathematical autonomy of occurrence and 
	bounded severity estimation.
	
	\vspace{10ex}
	
	\begin{table}[htbp]
		\centering
		\small
		\setlength{\tabcolsep}{5pt}
		\caption{Structural Comparison of Semicontinuous Modeling Paradigms}
		\label{tab:paradigm_comparison}
		\begin{tabularx}{\textwidth}{l >{\raggedright\arraybackslash}X >{\raggedright\arraybackslash}X >{\raggedright\arraybackslash}X}
			\toprule
			\textbf{Dimension} & \textbf{Conventional Hurdle} & \textbf{Tweedie Deviance} & \textbf{Proposed Latent ZIG} \\
			\midrule
			\textbf{Covariate Space ($\mathbf{X}$)} & Disjoint (full sample vs. positive subset) & Shared (single unified design matrix) & \textbf{Shared} (coherent feature representation) \\
			\addlinespace
			\textbf{Objective Function} & Fragmented (two separate deviance) & Unified scalar deviance & \textbf{Unified} scalar deviance ($\mathcal{D} = \mathcal{D}_{\text{occ}} + \mathcal{D}_{\text{sev}}$) \\
			\addlinespace
			\textbf{Parameter Coupling} & Decoupled, prone to discordant risk sorting & Rigidly coupled, $\mathrm{Var} = \phi \mu^p$, distorting zero mass & \textbf{Information-Orthogonal} ($\frac{\partial^2 \mathcal{D}}{\partial \zeta \partial \eta} \equiv 0$), zero distortion \\
			\addlinespace
			\textbf{Boundary Handling} & Unbounded or heuristic target clamping & Intractable series integration for $S(U)$ & \textbf{Closed-Form} incomplete Gamma survival accumulator \\
			\bottomrule
		\end{tabularx}
	\end{table}   
	\vspace{5ex}
	
	\subsection{The Tweedie Exponential Dispersion Family: Impact, Strengths, and Parameter Rigidity} \label{sec:tweedie_limits}
	
	The compound Poisson-Gamma Tweedie distribution \citep{Tweedie1984, Jorgensen1987} represents a foundational milestone in empirical risk analysis and predictive analytics. By capturing both an exact point mass at the origin and continuous positive outcomes within a single exponential dispersion structure, it provides an analytically parsimonious vehicle for pure premium estimation and property-casualty ratemaking \citep{Smyth2002, Dunn2018}. Its operational widespread adoption stems from practical convenience: it operates cleanly within Generalized Linear Model (GLM) routines, incorporates into gradient-boosted decision trees (GBDTs), and eliminates the administrative complexity of maintaining decoupled frequency and severity pipelines.
	
	Furthermore, canonical applications frequently hold the power parameter fixed at $p = 1.5$—the symmetric midpoint between Poisson counts ($p = 1$) and Gamma severity ($p = 2$)—which often delivers stable aggregate mean predictions on standard personal and small commercial portfolios without iterative profiling over $p$ \citet{Smyth2002}. 
	
	\noindent\textbf{Computational Profile: Deviance Optimization vs. Likelihood Evaluation}

	Despite its practical strengths, the Tweedie architecture exhibits a substantial computational asymmetry between regression scoring and density evaluation. For a predetermined variance power $p \in (1, 2)$, the unit deviance loss,
	\begin{equation}
		\mathcal{D}_{u}(y_i, \mu_i) = 2 \left( \frac{y_i^{2-p}}{(1-p)(2-p)} - \frac{y_i \mu_i^{1-p}}{1-p} + \frac{\mu_i^{2-p}}{2-p}\right), \quad \text{for } y_i \geq 0, \label{eq:unit_dev}
	\end{equation}
	evaluates via elementary closed-form operations, yielding clean score vectors and Hessians that scale efficiently in tree-based gradient optimization.
	
	Conversely, computing the full continuous density function or recovering exact survival probabilities requires evaluating an infinite series over latent Poisson frequency terms \citep{Dunn2005, Dunn2018}:
	
	\begin{equation}
		f_Y(y; \mu_i, \phi, p) = \frac{1}{y} \sum_{j=1}^{\infty} W_j(y, \phi, p), \quad y > 0. \label{eq:twd_series}
	\end{equation}
	Evaluating Eq.~\eqref{eq:twd_series} across large insurance portfolios demands computationally intensive series acceleration and Fourier inversion techniques to avoid floating-point breakdown. Moreover, when outcomes are subject to a deterministic capping boundary $U$ to curtail catastrophic leverage, evaluating the survival mass $S_{Y}(U) = \Pr(Y_{i} \ge U)$ requires numerical integration over this infinite series, making exact boundary score updates intractable during recursive tree splitting.  
	
	\noindent\textbf{Extended Quasi-Likelihood Estimation and the Deviance Post-Adjustment}

\begin{figure}[t]
		\centering
		\includegraphics[width=\linewidth]{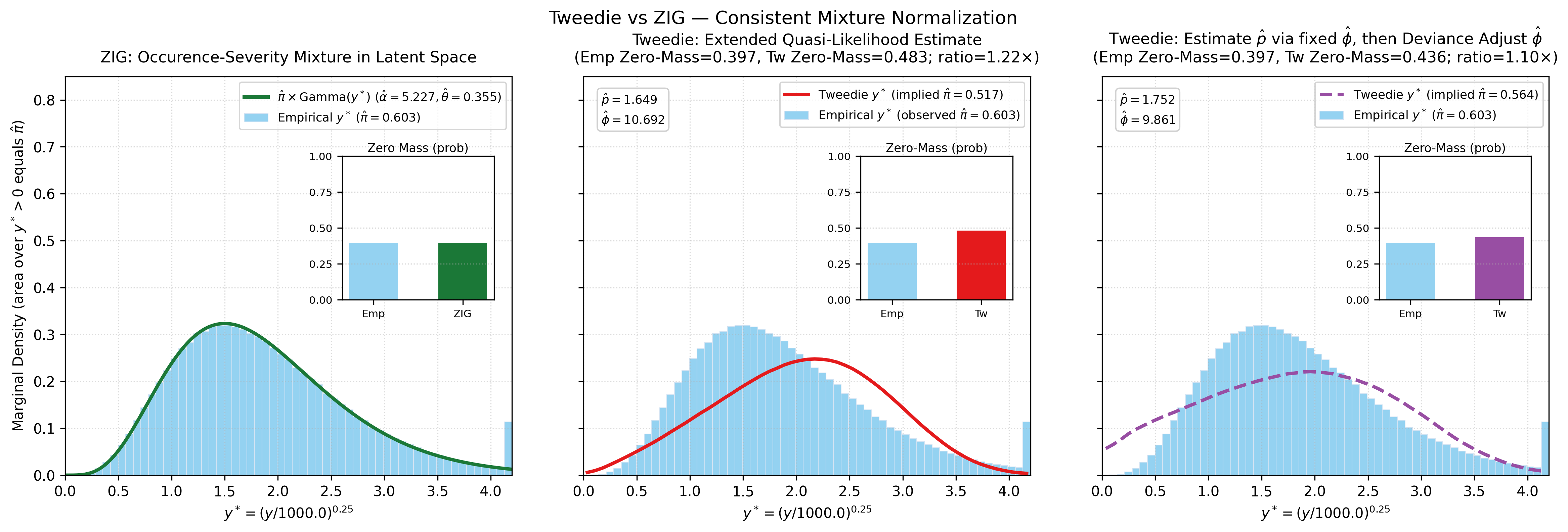}
		\caption{Empirical semicontinuous fits on a high-severity portfolio: the proposed latent ZIG framework (left) capturing both zero mass and continuous modal density, alongside the joint two-index Tweedie fit (center) and deviance-adjusted Tweedie fit (right) exhibiting zero-mass inflation and continuous body flattening.}
		\label{fig:tweedie_vs_zig}
	\end{figure}

	Because full maximum likelihood estimation of $\{p, \phi\}$ through infinite series summation is computationally prohibitive across large-scale datasets, Nelder and Pregibon’s Extended Quasi-Likelihood (EQL) \citep{Nelder1987} is standardly employed. For a sample of $n$ observations containing $n_{0}$ structural zeros and $n_{+}$ strictly positive outcomes ($n = n_{0} + n_{+}$), the negative profile log-EQL is:
	\begin{equation}
		-\ell_{\text{EQL}}(p, \phi) = \frac{1}{2} \sum_{i \in \{y_i > 0\}} \ln\left(2\pi \phi y_i^p\right) + \frac{1}{2\phi} \sum_{i=1}^n \mathcal{D}_{u}(y_i, \hat{\mu}_i)
		\label{eq:eql_objective}
	\end{equation}
	Optimizing the parameters exposes two primary calibration paradigms:Joint Two-Index Estimation $\{p, \phi\}$: Minimizing Eq.~\eqref{eq:eql_objective} jointly over $p \in (1.05, 1.95)$ balances the logarithmic normalization penalty against the global sample deviance $\mathcal{D} = \sum_{i=1}^{n} \mathcal{D}_{u}(y_{i}, \hat{\mu}_{i})$. As shown in the center pane of \figurename~\ref{fig:tweedie_vs_zig}, profile EQL yields $\hat{p} = 1.649$ and $\hat{\phi} = 10.692$.   One-Index Approach with Post-Hoc Deviance Adjustment: Alternatively, practitioners frequently fix the initial dispersion index (e.g., initialized at $\phi^{(0)} = 1.5$) to estimate the index parameter, yielding $\hat{p} = 1.752$. Subsequent deviance re-estimation across the sample drives the adjusted dispersion parameter to $\hat{\phi} = 9.861$ (\figurename~\ref{fig:tweedie_vs_zig}, right pane).   In both workflows, the estimated dispersion scale expands dramatically ($\hat{\phi} \approx 9.9 - 10.7$) to absorb the empirical sample variance, far exceeding the initial operational assumption of $\phi = 1.5$. Because the underlying Poisson frequency parameter scales inversely with dispersion,
	\begin{equation}
		\nu_{i} = \frac{\mu_{i}^{2-p}}{\phi (2 - p)}, \label{eq:zero_poison} 
	\end{equation}
	this massive inflation directly alters the zero-occurrence boundary.
	
	\noindent\textbf{The Coupled Parameter Paradox and Extreme-Tail Distortion}

	The fundamental limitation of the compound Poisson-Gamma Tweedie family lies in the mean-variance coupling that binds occurrence frequency to severity dispersion. In the Tweedie formulation, the discrete zero-mass probability is governed strictly by the zero-claim Poisson probability:
	\begin{equation}
		\Pr(Y_i = 0) = \exp(-\nu_{i}) = \exp\left( -\frac{\mu_i^{2-p}}{\phi (2 - p)} \right).
		\label{eq:tweedie_p0}
	\end{equation}
	Because global variance scales as $\mathrm{Var}(Y_{i}) = \phi \mu_{i}^{p}$, empirical portfolios exhibiting positive severities that span multiple orders of magnitude exert extreme variance demands on the estimator. Under joint likelihood or quasi-deviance minimization, absorbing this tail dispersion forces $\phi$ upward. However, because $\phi$ resides in the denominator of the exponent in Eq.~\eqref{eq:tweedie_p0}, an inflated $\phi$ drives the exponent toward zero, which artificially pushes the implied zero-mass probability $\Pr(Y_{i} = 0)$ upward toward 1.   
	
	To maintain aggregate sample calibration, the joint model is forced into a structural compromise: it over-allocates probability mass to the origin while flattening and over-dispersing the continuous density across the body and upper deciles. In production risk-segmentation environments, this artificial flattening severely impairs the model's granular sorting ability (e.g., Double Lift or Top-Decile Gains), muting distinctions between moderate losses and extreme tail claims.
	
	\noindent\textbf{Empirical Calibration and Diagnostic Positioning (\figurename~\ref{fig:tweedie_vs_zig})}

	To illustrate this structural distortion empirically, \figurename~\ref{fig:tweedie_vs_zig} benchmarks the Tweedie estimation approaches against the proposed latent-space Zero-Inflated Gamma (ZIG) formulation on a high-severity empirical portfolio.

	\vspace{5ex}
	\begin{table}[htbp]
		\centering
		\small 
		\setlength{\tabcolsep}{4.5pt} 
		\caption{Diagnostic Comparison across Latent ZIG and Tweedie Calibration Approaches}
		\label{tab:diagnostic_comparison}
		\begin{tabular}{lccc}
			\toprule 
			\textbf{Metric/Parameter} & \textbf{ZIG (Latent)} & \textbf{Twd (2-Index)} & \textbf{Twd (1-Index+Dev.)} \\
			\midrule  
			Calibration & $\hat\alpha = 5.227,\; \hat\theta = 0.355$ & $\hat{p} = 1.649,\; \hat{\phi} = 10.692$ & $\hat{p} = 1.752,\; \hat{\phi} = 9.861$ \\
			Obs. Zero Mass & 39.7\% & 39.7\% & 39.7\% \\
			Model Zero Mass & 39.7\% (exact match) & 48.3\% (ratio = $1.22\times$) & 43.6\% (ratio = $1.10\times$) \\
			Model Positive Prob. & 0.603 & 0.517 & 0.564 \\
			Tail/Body Fidelity & Mode Match & Flattened/Compressed & Flattened/Underfit \\
			\bottomrule  
		\end{tabular}
	\end{table}

	\vspace{5ex}
	
	\begin{itemize}
		\item Two-Index EQL Fit (Figure~\ref{fig:tweedie_vs_zig}, Middle Pane): Jointly optimizing $\{p, \phi\}$ produces $\hat{p} = 1.649$ and $\hat{\phi} = 10.692$. The excessive dispersion inflates the implied zero-mass to $48.3\%$ ($1.22\times$ the empirical zero occurrence), depressing the model's implied positive rate to $\hat{\pi} = 0.517$. Across the continuous domain ($y^* > 0$), the implied density (solid red curve) flattens severely, displacing the empirical mode and failing to capture positive-body concentration.   
		\item One-Index Deviance Fit (Figure~\ref{fig:tweedie_vs_zig}, Right Pane): Initializing at a standard $\phi^{(0)} = 1.5$ yields $\hat{p} = 1.752$, but the subsequent deviance-adjusted scale parameter expands to $\hat{\phi} = 9.861$. Although the zero-mass distortion is partially reduced ($43.6\%$ implied zeros, or $1.10\times$ empirical), the continuous density (dashed purple curve) remains substantially depressed across the modal region, overestimating extreme tail dispersion at the expense of body calibration. 
		\item  Dual-regime ZIG Framework (Figure~\ref{fig:tweedie_vs_zig}, Left Pane): Operating in the latent coordinate space $Y^* = (Y/s)^\lambda$, the ZIG model breaks this coupling entirely. An independent Bernoulli hurdle preserves the observed $39.7\%$ zero boundary identically ($\hat{\pi} = 0.603$). Concurrently, the regularized severity layer ($\alpha = 5.227, \theta = 0.355$) aligns with the continuous empirical support without structural feedback, capturing both body density and tail mass faithfully.
	\end{itemize}

	\section{The Scaled Power Transformation and Distributional Regularization}  \label{sec:power_trans}
	
	In semicontinuous modeling environments, positive continuous outcomes \(Y_i > 0\) frequently span multiple orders of magnitude, manifesting an extreme right tail that challenges standard parametric specifications. Rather than resorting to conventional logarithmic transformations... we utilize a specialized power transformation as given in Eq.~\eqref{eq:latent_mapping} to systematically eliminate distributional mismatch by controlling the empirical skewness of the raw positive continuous response variable, where \(s\) act as a scale-preserving adjustment factor. The regularized response variable \(Y_i^{*}\) is explicitly mapped to a much more compact support with familiar distributional shape.
	
	\subsection{Geometric Mechanics of the Scale Anchor} \label{sec:geo_mechanics}  
	While the power parameter $\lambda \in (0,1)$ governs distributional skewness, the reference scale $s > 0$ functions as an empirical pivot that defines the transition boundary between tail compression and lower-body expansion under the concave mapping $h(y) = (y/s)^\lambda$. The coordinate transformation bifurcates the positive support around the neutral unit anchor $y_i^* = 1$:
	\begin{align}
		y_i > s &\implies \left(\frac{y_i}{s}\right)^\lambda < \frac{y_i}{s} \quad \text{(concave right-tail compression)}, \label{eq:geometr1} \\
		0 < y_i < s &\implies \left(\frac{y_i}{s}\right)^\lambda > \frac{y_i}{s} \quad \text{(convex left-body expansion)}. \label{eq:geometr2}
	\end{align}
	In practice, the reference scale  $s \in (0, U)$ is selected prior to model estimation using robust empirical summaries of the positive support, such as the median, selected quantiles,  or trimmed means, informed both by distributional characteristics and domain-specific business considerations. This anchors the center of mass near unity while allowing $\lambda$ and $\alpha$ to govern tail curvature and dispersion. Furthermore, scaling by $s$ renders the ratio $(y_i/s)$ dimensionless, ensuring that the numerical arguments passed to the incomplete Gamma evaluations remain bounded within stable, overflow-free domains.
	
	\subsection{Statistical Properties of the Latent \texorpdfstring{\(\bm{Y}^*\)}{Y*}  Support} \label{sec:ytran_property}
	
	\begin{itemize}
		\item Decoupling Variance-Mean Hard-coding: Traditional modeling configurations (such as the standard Tweedie family) impose rigid exponential dispersion constraints where variance strictly tracks a power function of the mean, \(\text{Var}(Y_i^*) = \phi \mu_i^p\). Operating directly on the transformed support \(Y_i^{*}\) compresses the operational dynamic range of the heavy right tail prior to density estimation. This allows for flexible modeling of highly skewed processes without forcing an artificial parameter trade-off at the zero point-mass (\cite{Oshan2016}).
		\item Preserving Incomplete Gamma Argument Stability: Compressing the right-hand support using the scaling factor \(s\) and shape modifier \(\lambda \) maps extreme empirical realizations into a highly stable domain. This structural regularization prevents the numeric arguments passed into the lower incomplete Gamma function, \(\gamma(\alpha, z_i)\), from triggering numerical floating-point overflows or gradient blowouts across large-batch iterative optimization steps.
	\end{itemize}
	
	\section{Formulation and Optimization in the Latent Layer} \label{sec:formulation}
	
	All likelihood evaluations, expectation derivations, and loss curvature calculations operate directly within the latent coordinate space $(y_i^*, U^*)$ defined in Eq.~\eqref{eq:latent_mapping}.
	
	\subsection{Latent Space Hurdle Architecture and Distributional Pillars} \label{sec:three_pillars}
	
	Because the concave power transform $h(y) = (y/s)^\lambda$ is strictly monotonic on $[0, \infty)$ with $h(0) = 0$, occurrence probabilities are invariant between the original and latent spaces:
	\begin{align}
		\Pr(Y_i > 0 \mid \mathbf{x}_i) &= \Pr(Y_i^* > 0 \mid \mathbf{x}_i) = \pi(\mathbf{x}_i), \label{eq:occur1} \\[1ex]
		\Pr(Y_i = 0 \mid \mathbf{x}_i) &= \Pr(Y_i^* = 0 \mid \mathbf{x}_i) = 1 - \pi(\mathbf{x}_i). \label{eq:occur0}
	\end{align}
	
	Rather than coupling zero occurrence to severity dispersion, the semicontinuous outcome is decomposed into a Bernoulli hurdle and a bounded continuous severity law $G(y_i^* \mid \alpha, \beta_i)$:
	\begin{equation}
		f_{Y^*}(y_i^* \mid \mathbf{x}_i) = \big(1 - \pi(\mathbf{x}_i)\big) \, \mathbb{I}(y_i^* = 0) \;+\; \pi(\mathbf{x}_i) \, G(y_i^* \mid \alpha, \beta_i) \, \mathbb{I}(y_i^* > 0). \label{eq:two_segment_density}
	\end{equation}
	
	Conditioning on occurrence ($Y_i^* > 0$), the distribution of the positive outcome spans the remaining two physical pillars—the continuous Gamma modal body on $(0, U^*)$ and the accumulated survival mass at $U^*$—governed by the mixed density-survival law $G$:
	\begin{equation}
		G(y_i^* \mid \alpha, \beta_i) = 
		\begin{cases}
			\dfrac{\beta_i^\alpha}{\Gamma(\alpha)} (y_i^*)^{\alpha - 1} \exp\left(-\beta_i y_i^*\right), & 0 < y_i^* < U^*, \\[12pt]
			\dfrac{\Gamma(\alpha, \beta_i U^*)}{\Gamma(\alpha)}, & y_i^* \ge U^*,
		\end{cases} \label{eq:G_density}
	\end{equation}
	where $\alpha > 0$ is the global shape parameter, $\beta_i(\mathbf{x}_i) > 0$ is the covariate-dependent rate, and $\Gamma(\alpha, z) = \int_z^\infty t^{\alpha-1} e^{-t} \, dt$ is the upper incomplete Gamma function.

	\subsection{Closed-Form Expected Response in Original and Latent Spaces} \label{sec:closedform_expect}
	
	Because capping aggregates probability mass at $U^*$, the conditional expected severity in the latent layer differs from the unconstrained parameter. We evaluate the true conditional expectations directly under the positive law $G(y_i^* \mid \alpha, \beta_i)$.
	
	\paragraph{Latent Space Conditional Severity:}
	The conditional expected value of the bounded response $Y_i^* \mid Y_i^* > 0$ integrates across the continuous body and upper survival point mass:
	\begin{align}
		\mu_i^{*} &\equiv \mathbb{E}[Y_i^* \mid Y_i^* > 0, \mathbf{x}_i] \nonumber \\
		&= \int_0^{U^*} y^* \left[ \frac{\beta_i^\alpha}{\Gamma(\alpha)} (y^*)^{\alpha-1} e^{-\beta_i y^*} \right] dy^* \;+\; U^* \left[ \frac{\Gamma(\alpha, \beta_i U^*)}{\Gamma(\alpha)} \right] \nonumber \\
		&= \frac{1}{\beta_i} \left[ \frac{\gamma(\alpha + 1, \, z_i)}{\Gamma(\alpha)} + z_i \left(1 - \frac{\gamma(\alpha, z_i)}{\Gamma(\alpha)}\right) \right], \label{eq:cond_mu_star}
	\end{align}
	where $z_i = \beta_i U^* $ is the dimensionless boundary coordinate and $\gamma(\alpha, z) = \int_0^z t^{\alpha-1}e^{-t}dt$ is the lower incomplete Gamma function. Utilizing the property \( \gamma(\alpha + 1, \, z) = \alpha \, \gamma(\alpha, \, z) - z^\alpha e^{-z} \), Eq.~\eqref{eq:cond_mu_star} can be rewritten as
	\begin{equation}
		\beta_i \, \mu_i^{*} = \frac{(\alpha - z_i)\, \gamma(\alpha, \, z_i) -z_i^{\alpha}\, e^{-z_i}}{\Gamma(\alpha)} + z_i. \label{eq:beta_mu_star}
	\end{equation}
	
	The unconditional expected response in the latent layer is therefore:
	\begin{equation}
		\mathbb{E}[Y_i^* \mid \mathbf{x}_i] = \pi(\mathbf{x}_i) \, \mu_i^{*}(\mathbf{x}_i). \label{eq:expected_latent_uncond}
	\end{equation}
	It is important to note that the rate parameter $\beta_i$ of the mixed density-survival Eq.\eqref{eq:G_density} is governed by Eq.\eqref{eq:cond_mu_star}, with unique implicit inverse function 
	 \begin{equation}
	 	\beta_i = \psi(\mu_i^{*}\left(\mathbf{x}_{i}\right); \alpha, U^*). \label{eq:beta_rate}
	 \end{equation} 
	mapping the conditional expected response back to the rate parameter.
	
	\paragraph{Original Scale Conditional Severity:}
	Inverting the power transformation $y_i = s(y_i^*)^{1/\lambda}$, the conditional expected severity on the raw bounded scale $(0, U]$ is:
	\begin{align}
		\mu_i &\equiv \mathbb{E}[Y_i \mid Y_i > 0, \mathbf{x}_i] \nonumber \\
		&= \int_0^{U^*} s(y^*)^{1/\lambda} \left[ \frac{\beta_i^\alpha}{\Gamma(\alpha)} (y^*)^{\alpha-1} e^{-\beta_i y^*} \right] dy^* \;+\; U \left[ \frac{\Gamma(\alpha, \beta_i U^*)}{\Gamma(\alpha)} \right] \nonumber \\
		&= \frac{s}{\beta_i^{1/\lambda}} \frac{\gamma\left(\alpha + 1/\lambda, \, z_i\right)}{\Gamma(\alpha)} + U \left(1 - \frac{\gamma(\alpha, \, z_i)}{\Gamma(\alpha)}\right). \label{eq:cond_mu_raw}
	\end{align}
	Consequently, the unconditional expected payout (pure premium) on the raw operational scale evaluates in closed form:
	\begin{equation}
		\mathbb{E}[Y_i \mid \mathbf{x}_i] = \pi(\mathbf{x}_i) \, \mu_i(\mathbf{x}_i) = \pi(\mathbf{x}_i) \left[ \frac{s}{\beta_i^{1/\lambda}} \frac{\gamma\left(\alpha + 1/\lambda, \, z_i\right)}{\Gamma(\alpha)} + U \left(1 - \frac{\gamma(\alpha, z_i)}{\Gamma(\alpha)}\right) \right]. \label{eq:expected_response}
	\end{equation}
	
	\subsection{Link Parametrization and Joint Scalar Deviance Formulation} \label{sec:link_deviance}
	
	To operationalize the dual-regime architecture within high-dimensional learning engines (such as gradient-boosted decision tree ensembles), we map the structural distribution parameters to a vector of unconstrained additive link predictors $(\zeta_i, \eta_i)^T \in \mathbb{R}^2$ over a single, shared covariate matrix $\mathbf{X}$ \citep{Frees2010}. The predictive head maps the structural outputs through clear, unambiguous link mappings:
	\begin{equation}
		f_{\bm{\theta}}(\mathbf{x}_i) = 
		\begin{pmatrix} 
			\zeta_i \\ 
			\eta_i 
		\end{pmatrix} 
		= 
		\begin{pmatrix} 
			\logit(\pi_i) \\ 
			\log(\mu_i^{*}) 
		\end{pmatrix}
		=
		\begin{pmatrix}
			\log\left(\frac{\pi_i}{1 - \pi_i}\right) \\
			\log\left(\mu_i^{*}\right)
		\end{pmatrix}, \label{eq:link_mapping}
	\end{equation}
	where $\zeta_i$ governs the log-odds of positive occurrence, and $\eta_i$ establishes the log-scale of the true conditional expected latent response $\mu_i^{*}$. 
	
	By exploiting the monotonic property of the coordinate mapping $Y_i^* = (Y_i/s)^\lambda$, the sample log-likelihood decomposes into a strictly additive, two-part optimization task over $n$ independent observations:
	\begin{equation}
		\ell(\bm{\theta}; \bm{Y}^*) = \ell_{\text{occ}}(\bm{\zeta}; \bm{Y}^*) + \ell_{\text{sev}}(\bm{\eta}; \bm{Y}^* \mid \alpha),
	\end{equation}
	where the Bernoulli occurrence log-likelihood treats structural zeros as the reference class,
	\begin{equation}
		\ell_{\text{occ}}(\bm{\zeta}; \bm{Y}^*) = \sum_{i=1}^n \Big[ (1 - d_i)\log(1 - \pi_i) + d_i \log \pi_i \Big], \label{eq:loglik_occ}
	\end{equation}
	and the regularized severity log-likelihood evaluates strictly over the filtered positive sub-sample ($d_i = 1$). Keeping in mind the mapping Eq.~\eqref{eq:beta_rate}, the severity log-likelihood is expressed explicitly as:
	\begin{equation}
		\ell_{\text{sev}}(\bm{\eta}; \bm{Y}^* \mid \alpha) = \sum_{i: d_i = 1} \left[ (1 - \delta_i) \Big( \alpha \log \beta_i - \log\Gamma(\alpha) + (\alpha - 1)\log y_i^* - \beta_i y_i^* \Big) + \delta_i \log\left(\frac{\Gamma(\alpha, z_i )}{\Gamma(\alpha)}\right) \right], \label{eq:loglik_sev}
	\end{equation}
	utilizing the occurrence indicator $d_i = \mathbb{I}(y_i^* > 0)$ and upper boundary indicator $\delta_i = \mathbb{I}(y_i^* = U^*)$.
	
	To transform these log-likelihoods into a robust non-negative objective function minimized by gradient algorithms, we construct the total joint deviance 
	\begin{equation}
		\mathcal{D}(\bm{Y}^*; \bm{\theta}) = 2\Big[\ell\big(\tilde{\bm{\theta}}; \bm{Y}^*\big) - \ell\big(\bm{\theta}; \bm{Y}^*\big)\Big]  \label{eq:def_deviance}
	\end{equation}
	 evaluated against an unconstrained, fully saturated baseline model $\tilde{\bm{\theta}}$, where the parameters perfectly interpolate the observed conditional realizations ($\tilde{\pi}_i = d_i$ and $\tilde{\mu}_i^* = y_i^*$). 
	
	\paragraph{Saturated Baseline Parameter Estimators:}
	In the saturated model, the estimated conditional expectation matches the observed realization identically across all observations:
	\begin{enumerate}
		\item \textbf{Occurrence Layer ($y_i^* = 0$):} The saturated occurrence probability sets $\tilde{\pi}_i = 0$ and the corresponding indicator $d_i=0$, driving the zero log-likelihood contribution to its upper bound $\log(1 - \tilde{\pi}_i) = 0$.
		\item \textbf{Positive Severity Layer ($y_i^* > 0$):} The saturated conditional expectation is equated directly to the observed positive realization:
		\begin{equation}
			\tilde{\mu}_i^{*} \equiv  y_i^* \quad \forall \; i \text{ with } y_i^* > 0 \label{eq:saturated_identity}
		\end{equation}
		Implicit saturated rate parameter of Eq.~\eqref{eq:beta_rate} becomes
		\begin{equation}
			\tilde{\beta}_i = \psi(y_i^{*}; \alpha, U^*). \label{eq:beta_sat}
		\end{equation}
	\end{enumerate}
	
	Substituting these baseline estimators into the deviance framework yields the unified closed-form scalar objective function:
	\begin{equation}
		\mathcal{D}(\bm{Y}^*; \bm{\theta}) = \mathcal{D}_{\text{occ}}(\bm{Y}^*; \bm{\zeta}) + \mathcal{D}_{\text{sev}}(\bm{Y}^*; \bm{\eta} \mid \alpha),
	\end{equation}
	which expands to:
	\begin{align}
		\mathcal{D}_{\text{occ}}(\bm{Y}^*; \bm{\zeta}) &= -2 \sum_{i=1}^n \Big[ (1 - d_i)\log(1 - \pi_i) + d_i \log \pi_i \Big], \label{eq:final_dev_occ} \\[1ex]
		\mathcal{D}_{\text{sev}}(\bm{Y}^*; \bm{\eta} \mid \alpha) = 2 \sum_{i: d_i = 1} \Bigg[ &(1 - \delta_i) \left( \alpha \log\left(\frac{\tilde{\beta}_i}{\beta_i}\right) + (\beta_i - \tilde{\beta}_i)\; y_i^{*} \right) + \delta_i \log \left( \frac{\Gamma(\alpha, \tilde{z}_i)}{\Gamma(\alpha, z_i)} \right) \Bigg], \label{eq:final_dev_sev}
	\end{align}
	where \( \tilde{z}_i = \tilde{\beta}_i U^{*} \) and $\beta_i$ is evaluated implicitly from the link predictor via $e^{\eta_i} = \mu_i^{*}$. Equations \eqref{eq:final_dev_occ} and \eqref{eq:final_dev_sev} compose the final unified scalar loss function, establishing the exact coordinate system upon which modern tree-building loops compute optimal split geometries.

	\section{First- and Second-Order Optimization Derivatives}
	
	In gradient-boosted decision tree ensembles and modern Newton-Raphson solvers, optimization is performed with respect to the additive link predictors:
	\begin{equation}
		\zeta_i = \logit(\pi_i) \iff \pi_i = \frac{1}{1 + e^{-\zeta_i}}, \quad \text{and} \quad \eta_i = \log(\mu_i^{*}) \iff \mu_i^{*} = e^{\eta_i}.
	\end{equation}
	Here we derive the analytical elements of the gradient vector (Score) and Hessian matrix with respect to $\zeta_i$ and $\eta_i$ for the deviance loss objective $\mathcal{D}$.
	
	\subsection{The Gradient Vector (Score Elements)}
	Differentiating the deviance function with respect to the link predictors yields:
	\begin{enumerate}
		\item \textbf{Occurrence Link Coordinate ($\zeta_i$):}
		\begin{equation}
			g_{\zeta, i} = \frac{\partial \mathcal{D}}{\partial \zeta_i}  = 2 \left(\pi_i - d_i\right).
		\end{equation}
		\item \textbf{Severity Link Coordinate ($\eta_i$):} 
		\begin{equation}
			g_{\eta, i} = \frac{\partial \mathcal{D}}{\partial \eta_i}  
				= 2\left[ (1 - \delta_i) \left( \alpha\, \frac{\partial \log\beta_i }{\partial\eta_i} - 
				\frac{\partial \beta_i}{\partial\eta_i} \, y_i^* \right) - 
				\delta_i \, \frac{\partial\log\Gamma(\alpha, z_i)}{\partial z_i} \frac{\partial z_i}{\partial\eta_i} \right].  \label{eq:grad_sev}
		\end{equation}
		The derivative of $\beta_i$ with respect to $\eta_i$ in the above equation can be obtained by differentiating both sides of Eq.~\eqref{eq:beta_mu_star} as follows
		\begin{equation*}
			\frac{\partial\beta_i}{\partial\eta_i} \, \mu_i^{*} + \beta_i \, \mu_i^{*} = \left\{ \frac{1}{\Gamma(\alpha)} \left[ (\alpha - z_i)\, \frac{\partial \gamma(\alpha, z_i)}{\partial z_i} - \gamma(\alpha, z_i) - \alpha z_i^{\alpha -1} e^{-z_i} + z_i^{\alpha} e^{-z_i} \right] + 1 \right\} \frac{\partial z_i}{\partial\eta_i}.
		\end{equation*}
		Applying the Leibniz rule to the lower incomplete gamma function \( \frac{\partial \gamma(\alpha, z)}{\partial z} = z^{\alpha - 1} e^{-z} \), the first term cancels the third and fourth terms in curl bracket resulting in \( \left(1 - \frac{\gamma(\alpha, z_i)}{\Gamma(\alpha)} \right)  \). Recalling $z_i = \beta_i U^*$, it is easy to see \( \frac{\partial z_i}{\partial \eta_i} = U^* \frac{\partial\beta_i}{\partial\eta_i}\).
		The first order derivative can be deduced from the above equation: 
		\begin{equation}
			\frac{\partial \beta_i}{\partial \eta_i} = - \frac{\beta_i}{1  - \frac{U^*}{\mu_i^{*}} \left( 1 - \frac{\gamma(\alpha, z_i)}{\Gamma(\alpha)} \right) }.  \label{eq:beta_derivative}
		\end{equation}
		Substituting this result and  \( \frac{\partial\Gamma(\alpha, z)}{\partial z} = -z^{\alpha} e^{-z} \) into Eq.~\eqref{eq:grad_sev}, the exact severity gradient becomes:
		\begin{equation}
			g_{\eta, i} = \frac{2  \, \big[ \left(1 - \delta_i\right) \left( \alpha - \beta_i y_i^* \right) - \delta_i \, R(\alpha, z_i)\big]}{1 - \frac{U^*}{\mu_i^{*}} \left( 1 - \frac{\gamma(\alpha, z_i)}{\Gamma(\alpha)} \right) },  \label{eq:gradient_sev}
		\end{equation}
		where 
		\begin{equation}
			R(\alpha, z_i) \equiv \frac{z_i^\alpha e^{-z_i}}{\Gamma(\alpha, z_i)}
		\end{equation}
		denotes the regularized hazard component.
		
	\end{enumerate}
	
	\subsection{The Hessian Matrix Elements (Curvature)}
	The second partial derivatives define the local quadratic curvature of the loss surface:
	\begin{enumerate}
		\item \textbf{Occurrence Curvature ($\zeta_i$):}
		\begin{equation}
			h_{\zeta, i} = \frac{\partial^2 \mathcal{D}}{\partial \zeta_i^2} = 2\pi_i(1 - \pi_i).
		\end{equation}
		
		\item \textbf{Severity Curvature ($\eta_i$):}
		To express the second-order severity derivative compactly, define the upper boundary survival probability $S_i$, the implicit Jacobian denominator $D_i$, and the unscaled gradient numerator $N_i$:
		\begin{align}
			S_i &\equiv \frac{\Gamma(\alpha, z_i)}{\Gamma(\alpha)} = 1 - \frac{\gamma(\alpha, z_i)}{\Gamma(\alpha)}, \\
			D_i &\equiv 1 - \frac{U^*}{\mu_i^{*}} S_i, \\
			N_i &\equiv (1 - \delta_i)(\alpha - \beta_i y_i^*) - \delta_i R(\alpha, z_i),
		\end{align}
		such that the first-order severity gradient in Eq.~\eqref{eq:gradient_sev} takes the compact quotient form $g_{\eta, i} = 2 N_i / D_i$. Note that because $\mu_i^* > U^* S_i$ for any finite $\beta_i > 0$ (as established by Eq.~\eqref{eq:cond_mu_star}), the Jacobian factor satisfies $D_i \in (0, 1]$ strictly.
		
		Recalling $\partial z_i / \partial \eta_i = U^* (\partial \beta_i / \partial \eta_i) = -z_i / D_i$ from Eq.~\eqref{eq:beta_derivative}, the partial derivatives of the component terms with respect to $\eta_i$ evaluate cleanly:
		\begin{align}
			\frac{\partial R(\alpha, z_i)}{\partial \eta_i} &= R(\alpha, z_i)\big[\alpha - z_i + R(\alpha, z_i)\big] \frac{1}{z_i}\frac{\partial z_i}{\partial \eta_i} \nonumber \\
			&= -\frac{R(\alpha, z_i)\big[\alpha - z_i + R(\alpha, z_i)\big]}{D_i}, \label{eq:dR_deta} \\[1ex]
			\frac{\partial N_i}{\partial \eta_i} &= -(1 - \delta_i) y_i^* \frac{\partial \beta_i}{\partial \eta_i} - \delta_i \frac{\partial R(\alpha, z_i)}{\partial \eta_i} = \frac{h_{\eta,i}^{(0)}}{2 D_i}, \label{eq:dN_deta} \\[1ex] 
			h_{\eta,i}^{(0)} &\equiv 2\Big\{ (1 - \delta_i)\beta_i y_i^* + \delta_i R(\alpha, z_i)\big[\alpha - z_i + R(\alpha, z_i)\big] \Big\}, \label{eq:hess_0} \\[1ex]
			\frac{\partial D_i}{\partial \eta_i} &= \frac{U^*}{\mu_i^*} S_i - \frac{U^*}{\mu_i^*} \frac{\partial S_i}{\partial z_i}\frac{\partial z_i}{\partial \eta_i} \nonumber \\
			&= \frac{U^*}{\mu_i^*} S_i \left( 1 - \frac{R(\alpha, z_i)}{D_i} \right), \label{eq:dD_deta}
		\end{align}
		where $\partial S_i / \partial z_i = -z_i^{\alpha - 1} e^{-z_i} / \Gamma(\alpha) = -S_i R(\alpha, z_i) / z_i$ is applied in Eq.~\eqref{eq:dD_deta}. The definition \( h_{\eta,i}^{(0)} \) in Eq.~\eqref{eq:hess_0} corresponds to the unbounded Hessian component which will be shown later.
		
		Applying the quotient rule to $g_{\eta, i} = 2 N_i / D_i$ and substituting Eqs.~\eqref{eq:dN_deta}--\eqref{eq:dD_deta} yields the exact closed-form severity Hessian element:
		\begin{equation}
			h_{\eta, i} = \frac{\partial^2 \mathcal{D}}{\partial \eta_i^2} = 2 \left( \frac{1}{D_i}\frac{\partial N_i}{\partial \eta_i} - \frac{N_i}{D_i^2}\frac{\partial D_i}{\partial \eta_i} \right) 
			= \frac{1}{D_i^2} \left\{ h_{\eta,i}^{(0)} - g_{\eta, i} \frac{U^*}{\mu_i^*} S_i \big[ D_i  - R(\alpha, z_i) \big] \right\}. \label{eq:hessian_sev}
		\end{equation}
	\end{enumerate}
	If the gamma distribution is unbounded, the survival function $S_i$ = 0, \( \delta_i = 0 \), \( D_i = 1 \) and Eq.~\eqref{eq:cond_mu_star} is reduced to the familiar \( \mu_i^* = \alpha / \beta_i \).

	\subsection{Parameter Orthogonality and Multi-Output Gradient Coherence}
	Because the deviance decomposes additively into coordinate-separated link domains, 
	the cross-partial derivative is identically zero:
	\begin{equation}
		\frac{\partial^2 \mathcal{D}}{\partial \zeta_i \, \partial \eta_i} = 0.
	\end{equation}
	The global Hessian matrix $\bm{H}$ is strictly block-diagonal across all observations. 
	This property provides a profound computational advantage in multi-output tree boosting: 
	while the tree ensemble branches on a \textbf{single, shared covariate matrix} $\mathbf{X}$ to learn 
	coherent split structures, the leaf-weight updates for occurrence ($\Delta \zeta$) and bounded 
	severity ($\Delta \eta$) execute concurrently in closed form with zero cross-target gradient 
	contamination or alternating optimization delay.

	\section{Boundary Stability and Asymptotic Regularization Analysis}

	The numerical stability of the optimization loop depends on the asymptotic behavior of $R(\alpha, z_i)$ and the implicit Jacobian factor $D_i$ at the upper boundary $\delta_i = 1$ as the dimensionless coordinate $z_i = \beta_i U^*$ approaches extreme values.
	
	\subsection{Deep-Tail Asymptotic Boundedness}
	When a boundary observation ($\delta_i = 1$) exhibits a small predicted bounded latent mean $\mu_i^*$ relative to the capping threshold $U^*$ ($\mu_i^* \ll U^*$), the implicit rate $\beta_i = \psi(\mu_i^*; \alpha, U^*)$ grows large, driving the dimensionless argument $z_i = \beta_i U^*$ toward infinity ($z_i \to \infty$). Using the classical asymptotic expansion of the upper incomplete Gamma function,
	\begin{equation}
		\Gamma(\alpha, z) = z^{\alpha - 1} e^{-z} \left[ 1 + \frac{\alpha - 1}{z} + \mathcal{O}(z^{-2}) \right] \quad \text{as } z \to \infty,
	\end{equation}
	the regularized hazard component satisfies:
	\begin{equation}
		R(\alpha, z_i) = \frac{z_i^\alpha e^{-z_i}}{\Gamma(\alpha, z_i)} = \frac{z_i}{1 + \frac{\alpha - 1}{z_i} + \mathcal{O}(z_i^{-2})} = z_i - (\alpha - 1) + \mathcal{O}(z_i^{-1}).
	\end{equation}
	Furthermore, in this regime the boundary survival probability $S_i = \Gamma(\alpha, z_i)/\Gamma(\alpha) \sim z_i^{\alpha - 1} e^{-z_i} / \Gamma(\alpha)$ decays exponentially. From Eq.~\eqref{eq:cond_mu_star}, $\gamma(\alpha + 1, z_i)/\Gamma(\alpha) \to \alpha$ and $z_i S_i \to 0$, so the implicit rate converges to the uncapped relation $\beta_i \sim \alpha / \mu_i^*$, yielding $z_i \sim \alpha U^* / \mu_i^*$. Because $\frac{U^*}{\mu_i^*} S_i \sim \frac{z_i^\alpha e^{-z_i}}{\alpha \Gamma(\alpha)} \to 0$ exponentially fast, the Jacobian factor in Eq.~\eqref{eq:gradient_sev} approaches unity ($D_i \to 1$):
	\begin{equation}
		\lim_{z_i \to \infty} R(\alpha, z_i) \sim z_i \sim \frac{\alpha U^*}{\mu_i^*}, \qquad \lim_{z_i \to \infty} D_i = 1.
	\end{equation}
	
	Substituting these limiting forms into the severity gradient $g_{\eta, i}$ in Eq.~\eqref{eq:gradient_sev} and Hessian $h_{\eta, i}$ in Eq.~\eqref{eq:hessian_sev} at $\delta_i = 1$:
	\begin{align}
		\lim_{z_i \to \infty} g_{\eta, i} \Big|_{\delta_i = 1} &\approx -2 z_i \sim -2 \frac{\alpha U^*}{\mu_i^*}, \\[1ex]
		\lim_{z_i \to \infty} h_{\eta, i} \Big|_{\delta_i = 1} &\approx 2 z_i \sim 2 \frac{\alpha U^*}{\mu_i^*}.
	\end{align}
	
	This confirms two critical mathematical properties of the optimization engine:
	\begin{enumerate}
		\item \textbf{Asymptotic Positive Curvature:} For upper-boundary observations in the deep-underprediction regime  \( z_i \rightarrow \infty \), the severity Hessian satisfies \( h_{\eta,i} \sim 2 z_i > 0.\) Consequently, the Newton step remains well-conditioned. The Hessian need not be positive for all finite $z_i$, so the severity deviance is not globally convex in \( \eta_i \). This local variation in curvature does not, however, preclude the existence of a well-defined minimum of the deviance.
		\item \textbf{First-to-Second Order Balance:} The Newton step for an underestimated boundary point is bounded:
		\[ \Delta \eta_i = -\frac{g_{\eta, i}}{h_{\eta, i}} \longrightarrow -\frac{-2z_i}{2z_i} = +1 \]
		as the boundary slope $g_{\eta, i} \to -2z_i$ and the boundary curvature $h_{\eta, i} \to 2z_i$.
	\end{enumerate}
	
	Unlike unbounded gradient objectives or sigmoidal classifications where extreme tail outcomes suffer from vanishing gradients ($\sim 0$) or catastrophic divergence ($\sim \infty$), the boundary gradient scales linearly in $z_i$ with respect to the inverse latent mean. This linear scaling acts as an intrinsic quadratic restoring force, preventing gradient blowout while penalizing underestimation of high-risk boundary cases.

	\section{Global Parameter Estimation via Latent Profile Deviance} \label{sec:hyperparameters}
	
	The performance, numerical stability, and geometry of the proposed zero-inflated Gamma framework rely on a set of global structural parameters: the latent-space coordinate parameters $(\lambda, s)$ specified in Section~\ref{sec:power_trans}, alongside the continuous Gamma shape parameter $\alpha > 0$ introduced in Section~\ref{sec:formulation}. Rather than treating these parameters as heuristic algorithmic tuning knobs, we formalize their estimation through a concentrated profile deviance criterion evaluated directly on the latent coordinate space.
	
	\subsection{Latent Space Configuration and Bounded Likelihood Calibration} \label{sec:latent_config}
	
	The latent coordinate geometry is governed by the scale anchor $s > 0$ and power exponent $\lambda \in (0, 1)$, alongside the upper boundary $U^* = (U / s)^\lambda$. Following the variance-stabilizing and coordinate-centering criteria introduced in Section~\ref{sec:intro}, the scale anchor $s$ is chosen to match the order of magnitude of central severity (e.g., $s = 1\,000$), ensuring transformed outcomes remain centered near unity. The power parameter $\lambda$ is selected over a discrete candidate grid $\lambda \in \{0.15, 0.20, 0.25, 0.30\}$, balancing skewness attenuation against tail resolution and avoiding empirical boundary collapse.
	
	Conditional on a fixed coordinate transformation $(\lambda, s)$ and upper boundary $U^*$, the continuous shape parameter $\alpha$ is estimated directly from the positive support $\{y_i^* > 0\}$. To properly account for policy limit capping without distorting continuous dispersion, we calibrate $\alpha$ via profile maximum likelihood estimation on the right-censored Gamma law $G(y_i^* \mid \alpha, \beta_i)$ defined in Eq.~\eqref{eq:G_density}, where $\beta_i = \psi(\mu_i^*; \alpha, U^*)$ maps the bounded conditional mean $\mu_i^*$ to the underlying Gamma rate:
	\begin{equation}
		\hat{\alpha} = \arg\max_{\alpha > 0} \; \sum_{i: d_i = 1} \log G\big(y_i^* \;\big|\; \alpha, \psi(\mu_i^*; \alpha, U^*)\big).
	\end{equation}
	Evaluating $\hat{\alpha}$ under this boundary-adjusted formulation prevents the synthetic inflation of variance that typically arises when treating capped point masses as standard unbounded observations, ensuring that the empirical dispersion aligns consistently with the regularized deviance framework.
	
	To understand the profiling mechanism, it is instructive to view estimation through the spectrum of model complexity across both Generalized Linear Models (GLMs) and Gradient-Boosted Decision Trees (GBDTs). Both paradigms operate between two foundational theoretical extremes:
	\begin{enumerate}
		\item \textbf{The Lower Extreme (Null Model / Root Node):} In classical GLMs, the null model $\mathcal{M}_0$ constrains all slope coefficients to zero ($\bm{\beta} = \mathbf{0}$), estimating only the global intercept $\beta_0$ where unexplained deviance is maximized (the Null Deviance $\mathcal{D}_0$). In tree ensembles, this corresponds to the unsegmented root node $T_{\text{root}} = f_0(\mathbf{x})$, initialized uniformly at the global base margin before any feature splitting occurs.
		\item \textbf{The Upper Extreme (Saturated Baseline):} In GLMs, the saturated model $\mathcal{M}_{\text{sat}}$ assigns an individual parameter to every observation ($p = n$), interpolating each data point ($\tilde{\mu}_i^* = y_i^*$) such that deviance is identically zero ($\mathcal{D}_{\text{sat}} = 0$). In tree ensembles, this corresponds to an unconstrained tree $T_{\text{sat}}$ grown until every terminal leaf isolates a single observation, maximizing sampling variance.
	\end{enumerate}
	
	Optimal predictive estimation seeks an interior model---whether regularized GLM coefficients or an ensemble of depth-constrained trees $\sum_{m=1}^M f_m(\mathbf{x})$---that minimizes out-of-sample loss along the bias-variance frontier. However, because tree-splitting algorithms evaluate gradient steps $(g_i, h_i)$ derived from the objective function $\mathcal{D}$, the coordinate geometry governing the latent response cannot shift dynamically during tree construction without inducing optimization divergence.
	
	Consequently, the global structural parameters $(\lambda, s, \alpha)$ must be estimated at the unsegmented lower extreme ($\mathcal{M}_0 \equiv T_{\text{root}}$) prior to recursive feature space partitioning. Because the deviance function $\mathcal{D}(\bm{Y}^*; \bm{\theta})$ given in Eqs.~\eqref{eq:final_dev_occ}--\eqref{eq:final_dev_sev} exhibits complete parameter orthogonality between occurrence and severity, and because $\mu_i^*$ directly parameterizes the bounded conditional expectation $\mathbb{E}[Y_i^* \mid Y_i^* > 0]$, the intermediate nuisance parameters $\bm{\pi}$ and $\bm{\mu}^*$ at this root initialization state concentrate out directly to their empirical null moment estimators:
	\begin{equation}
		\hat{\pi} = \frac{1}{n} \sum_{i=1}^n d_i, \quad \text{and} \quad \hat{\mu}^*(\lambda, s) = \frac{1}{\sum_{i=1}^n d_i} \sum_{i: d_i = 1} y_i^*,
		\label{eq:null_estimators}
	\end{equation}
	with corresponding null rate $\hat{\beta}_0 = \psi(\hat{\mu}^*(\lambda, s); \alpha, U^*)$.
	
	Evaluating the deviance across the sample using these closed-form null initializers defines the concentrated profile deviance surface $\mathcal{D}_p(\alpha, \lambda, s; \bm{Y}^*)$ on the latent coordinate space:
	\begin{equation}
		\mathcal{D}_p(\alpha, \lambda, s; \bm{Y}^*) = \inf_{\bm{\pi}, \bm{\mu}^*} \mathcal{D}(\bm{Y}^*; \bm{\pi}, \bm{\mu}^*, \alpha) = \mathcal{D}\big(\bm{Y}^*; \hat{\bm{\pi}}, \hat{\bm{\mu}}^*(\lambda, s), \alpha\big).
	\end{equation}
	
	This fixed coordinate frame defines the invariant loss surface upon which subsequent tree-building engines partition the covariate space $\mathbf{x}$ to drive deviance reduction toward the optimal bias-variance equilibrium.
	
	\subsection{Variance Stabilization and the Latent Geometry Frontier} \label{sec:geom_frontier}
	
	The estimation of $\lambda$ operates along a critical statistical frontier balancing tail variance stabilization against resolution at the origin:
	\begin{itemize}
		\item \textbf{Under-regularization ($\lambda \to 1$):} Retaining a near-linear scale forces the continuous Gamma shape parameter $\alpha$ to collapse toward zero ($\alpha \to 0$) to accommodate extreme right-tail leverage. This collapse introduces numerical instability into the incomplete Gamma evaluations within the censored deviance component, triggering gradient blowouts.
		\item \textbf{Over-regularization ($\lambda \to 0$):} Compressing the support too aggressively toward a logarithmic boundary collapses the dynamic range of the tail, causing the continuous body to artificially bunch up near the zero threshold. This over-compression warps the local latent geometry, obscuring the risk-sorting features of the continuous support.
		\item \textbf{Balance of Left- and Right-Tail Geometry ($s \in (0,U)$):} Choosing $s$ too close to the capping boundary $U$ over-expands the lower body of the distribution relative to the anchor, whereas setting $s$ excessively small compresses the right tail too aggressively, artificially bunching the continuous density near the origin.
	\end{itemize}
	
	The estimated parameters (such as the empirically derived $\hat{\lambda} = 0.25$, $s = 1000$, and $\hat{\alpha} = 5.227$ manifested in our baseline calibration) represent the exact geometric equilibrium where the transformed positive support achieves a stabilized variance structure, enabling the continuous severity component to be modeled by a standard, stable Gamma density.
	
	\subsection{Algorithmic Estimation via Two-Stage Profile Optimization} \label{sec:two_stage_opt}
	
	To preserve computational efficiency across large datasets, the profile deviance minimization is operationalized via a performant two-stage protocol:
	\begin{enumerate}
		\item \textbf{Coarse Global Scan:} A grid search is evaluated across candidate values of $\lambda \in [0.05, 0.95]$ and $\alpha \in [0.5, 20.0]$, with $s$ evaluated over robust empirical anchor candidates (e.g., median or trimmed means of the positive support). At each grid coordinate, the latent profile deviance is evaluated via vectorized operations.
		\item \textbf{Local Newton--Raphson Refinement:} Once the neighborhood of the global minimum is identified, a continuous optimization routine is initialized. Because the profile surface is smoothly differentiable with respect to $\alpha$ and $\lambda$, a standard derivative-based solver converges rapidly to the analytical optimum.
	\end{enumerate}

	\section{Concluding Remarks and Discussion}
	
	This paper has established a unified, mathematically coherent likelihood framework for modeling semicontinuous, heavy-tailed data subject to explicit upper boundary constraints. Semicontinuous processes routinely challenge classical statistical models due to simultaneous, conflicting distributional demands: high zero-prevalence on the lower boundary and multi-order-of-magnitude dispersion in the positive continuous tail. Historically, practitioners were forced to choose between the operational simplicity of single-index Tweedie models---which compromise extreme-risk discrimination due to the parameter rigidity paradox---or decoupled hurdle models that lack tractable gradients when loss realizations are capped.
	
	The fundamental methodological shift introduced in this work is the strict confinement of statistical optimization to a regularized latent modeling space $\bm{Y}^*$ governed by the joint mapping as shown in Eq.~\eqref{eq:latent_mapping}.
	By scaling positive outcomes relative to an empirical pivot $s \in (0, U)$ and controlling curvature via $\lambda \in (0, 1)$, this coordinate mapping achieves global distributional regularization. Extreme positive realizations are compressed, smaller positive realizations are expanded, and the argument space of the lower incomplete Gamma function is confined within a stable, overflow-free numerical domain.
	
	Architecturally, the framework partitions the empirical support into three well-defined regimes:
	\begin{enumerate}
		\item \textbf{A degenerate zero point-mass} ($y_i^* = 0$) handled via an unconstrained Bernoulli occurrence layer.
		\item \textbf{A finite continuous positive body} ($0 < y_i^* < U^*$) governed by an analytical Gamma density.
		\item \textbf{A capped boundary point-mass} ($y_i^* = U^*$) governed by closed-form incomplete Gamma survival ratios.
	\end{enumerate}
	Crucially, the upper boundary threshold $U$ is not treated as a heuristic data-cleaning artifact or arbitrary outlier truncation. Instead, it serves as an intrinsic mathematical boundary where tail survival mass accumulates directly within the likelihood. This formulation eliminates the pathological numerical behavior that typically plagues heavy-tailed estimation: it prevents gradients from flattening, ensures the loss surface remains well-conditioned, and guarantees that a handful of extreme observations cannot exert disproportionate leverage over the optimization geometry.
	
	From an optimization perspective, the framework delivers five key computational properties:
	\begin{itemize}
		\item \textbf{Coherent Multi-Output Learning over Shared Feature Space:} Unlike traditional 
		hurdle approaches that bifurcate estimation across fragmented datasets and divergent 
		feature subsets, the proposed framework estimates occurrence probability and bounded severity 
		concurrently over a single, unified covariate matrix $\mathbf{X}$, guaranteeing aligned risk 
		segmentation across the entire portfolio.
		\item \textbf{Unified Additive Deviance:} Occurrence classification and bounded severity 
		regression are synthesized into a single scalar objective $\mathcal{D}(\bm{Y}^*; \bm{\theta})$ 
		evaluated against an exact saturated baseline.
		\item \textbf{Information Orthogonality without Cross-Talk:} The joint Hessian is strictly 
		block-diagonal ($\partial^2 \mathcal{D} / \partial \zeta_i \partial \eta_i \equiv 0$), allowing 
		multi-target tree leaves to update occurrence and severity simultaneously without numerical 
		feedback or stability degradation.
		\item \textbf{Deep-Tail Quadratic Safeguard:} In the deep right tail ($z_i \to \infty$), the score vector and Hessian scale symmetrically ($g_{\eta, i} \to -2z_i$, $h_{\eta, i} \to 2z_i$), bounding the Newton--Raphson step to $\Delta \eta_i \approx +1$. This guarantees strictly positive curvature ($h_{\eta, i} > 0$) and eliminates the vanishing-gradient pathologies common to sigmoidal or infinite-series formulations.
		\item \textbf{High-Performance Execution:} Because all scores and curvatures resolve to closed-form algebraic and incomplete Gamma operations, the loss engine executes in $\mathcal{O}(1)$ time per observation, lending itself naturally to vectorized CUDA and C++ kernels for multi-output gradient boosting.
	\end{itemize}
	
	While this manuscript formalizes the mathematical foundation and parametric behavior of the framework, its analytical design makes it an ideal drop-in loss engine for high-dimensional predictive algorithms. In our forthcoming companion applied manuscript, we transition from theoretical derivation to large-scale empirical execution. There, we demonstrate the operational performance of this framework implemented directly within GPU-accelerated XGBoost architectures, presenting rigorous evaluations of out-of-sample deviance stability, double-lift risk sorting, and dual Gains curves benchmarked against the theoretical Oracle Bound across multi-tier insurance portfolios.

	\bibliographystyle{plainnat}
	\bibliography{zig_ref}
	
\end{document}